\documentclass[11pt,prd,preprintnumbers,eqsecnum,superscriptaddress,nofootinbib]{revtex4}
\usepackage[usenames,dvipsnames]{xcolor}
\usepackage{graphics,epsfig}
\usepackage{graphicx}
\usepackage{color}
\usepackage{amssymb,amsmath,amsfonts}
 \usepackage{amsmath}
\usepackage{amssymb}
\usepackage{hyperref}
\usepackage{epstopdf}

\def\td#1{\tilde{#1}}
\def\check{ \maltese {\bf Check!}}

\begin{document}
\def\td#1{\tilde{#1}}
\def\check{ \maltese {\bf Check!}}

	\vspace{25mm}
	
	\begin{center}
		{\Large \bf Deuteron gravitational form factors, generalized parton distributions, and charge density in the framework of the soft-wall AdS/QCD model}
		
		\vskip 1. cm
		{Shahin Mamedov $^{a,b,c,e}$\footnote{e-mail : sh.mamedov62@gmail.com}, 
			Minaya Allahverdiyeva $^{b,c,d}$\footnote{e-mail : minaallahverdiyeva@ymail.com},
			and Narmin Akbarova $^{a,c}$\footnote{e-mail : nerminh236@gmail.com}}
		
		\vskip 0.5cm
		{\it $^a\,$ Institute for Physical Problems, Baku State University, Z.Khalilov 23, Baku, AZ 1148, Azerbaijan}\\
		{\it $^b\,$ Institute of Physics, Ministry of Science and Education, H.Javid 33, Baku, AZ 1143, Azerbaijan}\\
		{\it $^c\,$ Center for Theoretical Physics, Khazar University, 41 Mehseti Str., Baku, AZ1096, Azerbaijan}\\
            {\it $^d\,$ Western Caspian University, 31 Istiqlaliyyat Str., Baku, AZ1001, Azerbaijan}\\
            {\it $^e\,$ Lab. for Theor. Cosmology, International Centre of Gravity and Cosmos, Tomsk State \\ University
of Control Systems and Radio Electronics (TUSUR), 634050 Tomsk, Russia}
 
	\end{center}
	
	\centerline{\bf Abstract} \vskip 4mm

We study gravitational deuteron form factors (GFF) and generalized parton distributions (GPDs) within the soft-wall AdS/QCD model, where deuteron is described by the bulk vector field with twist $\tau=6$. For finite-temperature studies, we apply the soft-wall model, which is thermalized by introducing a thermal dilaton field. The GPDs and charge density are considered in impact parameter (IP) space and at zero and finite temperatures. We plotted the temperature dependence of these quantities in the IP space and observe a decrease in their peaks as the temperature increases. The gravitational root mean square radius obtained here is close to the range given by experimental data for the mass radius and has low sensitivity to the temperature. 
	
	\vspace{1cm}
	\section{Introduction}
	Studying the internal structure of hadrons including a deuteron is an important topic in nuclear physics. Deep virtual Compton scattering or deep virtual meson scattering experiments are good tools for studying hadrons.   The main theoretical objects in this study are the gravitational and electromagnetic form factors and related quantities. Other form factors such as axial-vector and pseudoscalar ones also provide valuable information for the structure investigation. The deuteron GFFs describe the interaction vertex in the graviton-deuteron elastic scattering and, being the form factors of the energy-momentum tensor (EMT), give the knowledge about the distribution of energy-momentum inside this nucleus. The GPDs determine the distribution of the partons inside a particle, depending on the fractional momentum of the parton and the skewness of the particle. These two quantities, i.e. GFFs and GPDs, are mathematically related. Experimentally, the GPDs are determined in the experiments at JLAB, Hall A, and COMPASS at CERN. Mathematically, they are found using the relations between GPDs and the form factors.  The deuteron GPDs were studied in the effective Lagrangian approach in Ref.~\cite{27} and employing GPDs the $G_C(Q^2)$, $G_M(Q^2)$, and $G_Q(Q^2)$ form factors were obtained. The GPDs appear in processes of deeply virtual Compton scattering (DVCS) or vector meson productions \cite{28,29,30,31} and they give the ordinary parton distributions in the forward limit. GPDs for the deuteron were introduced in Ref. \cite{32} and studied theoretically in Refs. \cite{33,34,35,36,37,38,39,40,41}.
    After the  AdS/QCD models were built, which are based on the AdS/CFT correspondence principle \cite{M, G, W}, the different physical quantities of hadron phenomenology were successfully studied within these models \cite{1,2,3,4,5,6,7,8,9,10, Rinaldi,11,12,13,14,17}. Seminal works devoted to the study of GFFs for vector and spinor particles are \cite{1,2,3}, where the study was performed within the hard-wall model of AdS/QCD and the relationship between GFFs and GPDs has also been shown.  The GFFs for vector meson, developed within the hard-wall model, were extended to the deuteron case within the soft-wall model in Ref. \cite{8}, where deuteron GPDs have been obtained from EM FFs. Electromagnetic form factors (EM FFs) of the deuteron were investigated within the soft-wall model in Refs. \cite{4,5} and within the hard-wall model in Ref. \cite{6}. Alternatively, the deuteron GPDs can be obtained from GFFs within the AdS/QCD framework. So, it is interesting to derive the GPDs from the GFFs and compare them with the ones obtained from the EM FFs. Several works have been devoted to the study of deuteron within the top-down approach of holographic QCD, as well (\cite{49,48,50,51}).

	Soft and hard-wall models have also been applied in the study of the strong interaction coupling constants and FFs of hadrons in the thermal medium \cite{18,19,20,21,22,23,24,25,26}. In Refs. \cite{19,20} by making the dilaton field thermal, the soft-wall model has been extended to the nonzero temperature case and the explicit profile functions for the free boson and fermion fields interacted with the thermal dilaton field were found. Using this model, the temperature dependence of the meson-nucleon coupling constants, the nucleon's axial-vector FFs, and the vector meson GFFs and GPDs were studied in the finite-temperature case in Refs. \cite{21,22,23,24,25,26}. In some physical situations the deuteron can turn out in the thermal medium, say in the Sun or heavy-ion collisions. So, changing its internal structure and properties depending on the temperature is interesting for studies in hot medium. The thermalized soft-wall model opens an opportunity for the study of the deuteron GFFs in the finite-temperature case as well. As a consequence, the temperature dependence of the gravitational radius of the deuteron can be determined using the thermal GFFs obtained in this model.  The GPDs for $\rho$ mesons were obtained from the GFFs within the soft-wall model at zero and finite-temperature cases in Ref. \cite{24} AdS/QCD. The soft-wall model opens wide opportunities for the study of deuteron GPDs at zero and finite temperatures \cite{24}.  
	
	For the unpolarized $\rho$ mesons, the charge density was derived from the GFFs within both hard and soft-wall models \cite{3}. The charge density for the transversely polarized deuteron \cite{8} was obtained from the EM FF within the soft-wall model, where a study of deuteron GFFs was also carried out. Following these works, the charge density for the unpolarized deuteron can be obtained from the GFFs within the holographic models. The finite-temperature soft-wall model opens an opportunity to study the temperature dependence of the deuteron charge density as well.
	
	The impact parameter (IP) shows the separation of the struck quark from the center of momentum, and the soft-wall study of the deuteron GPDs and charge density can be extended to the IP space. Within the soft-wall AdS/QCD model, the deuteron charge density was plotted in two-dimensional graphs in Ref. \cite{8}. The transverse charge density transversely polarized deuteron within the soft-wall model was studied in Ref. \cite{47}  It is possible to extend this plot to three dimensions, and finite-temperature cases as well.  
	
	The paper is organized in the following: In Section II, we present the basic elements of the soft-wall model and study the deuteron GFFs and gravitational radius at zero and finite-temperature cases. In Section III, we study deuteron GPDs obtained from the GFFs and consider their temperature dependence. In Section IV, we investigate deuteron GPDs in the IP space at zero and finite temperatures. In Section V, we calculate the deuteron charge density at zero and finite temperature. We summarize our results for deuteron GFFs, GPDs, and charge density in the last section.
		
	\bigskip
	\section{Deuteron GFFs at zero and finite temperatures} \label{GFFs}
	
	The background geometry for the model is the $5$-dimensional (5D) AdS space with the radius $R=1$, which is given by the following line element in Poincare coordinates:
	\begin{equation}
	ds^2=g_{MN}dx^{M}dx^{N}
	=\frac{1}{z^2}\left(-dz^2+(\eta_{\mu\nu}+h_{\mu\nu})dx^{\mu} dx^{\nu}\right),\quad \mu,\nu=0,1,2,3;\quad M,N=0,1,2,3,5.
	\label{1}
	\end{equation}
	Here $\eta_{\mu\nu}=diag(1,-1,-1,-1)$ is the 4-dimensional Minkowski metric and $z$ is the holographic coordinate varying in the range $\epsilon\leq z < \infty$. $h_{\mu\nu}$ is the perturbation of the metric and $h_{5N}=0$.
	
Finiteness of the 5D action is provided by introducing an exponential dilaton field $e^{-\Phi(z)}$ with $\Phi\left(z\right)=k^2 z^2$, which suppresses expressions under the integral over $z$ at infinity. This field also provides the dual QCD theory's confinement and chiral symmetry-breaking properties. For the $k$ parameter we take value $k=190 ~{\rm MeV}=0.9 ~{\rm fm}^{-1}$, which was applied in Refs. \cite{7,8} and fixed by fitting the electromagnetic form factors of the deuteron with experimental data.
 
 Action for the model of deuteron interacting with the graviton will be written as: 
	\begin{equation}
	S=\int_{0}\limits^{\infty}d^4x dz e^{-\Phi(z)} \sqrt{g} \left[L_{d}(x,z)+L_g(x,z)\right],
	\label{2}
	\end{equation}
	where $L_d(x,z)=- D^M d^\dagger_{N} D_M d^N +d^\dagger_{M} \, \Big(\mu^2 +  U(z) \Big) \, d^M$ is a deuteron lagrangian and 
	\begin{equation} \nonumber
	\mu^{2}=\left(\Delta-1\right)\left(\Delta-3\right)=\left(L+5\right)\left(L+3\right).
	\end{equation}
	Here $\Delta=\tau+L$ is the conformal dimension of the bulk vector field $d^{N} \left(x,z\right)$ describing the deuteron in the boundary theory, $L=max |L_z|$ is the maximal value of the $z$ component orbital angular momentum of the quark field.  $L_g(x,z)=R+12$ in the Eq (\ref{2}) is the lagrangian for gravity sector, $R$ is the Ricci scalar, $g=|\det g_{MN}|$. The potential $U(z)$ for the deuteron was found in Ref. \cite{5} and has the explicit form $U(z)=\phi(z) U_0$, where the value $U_0 = 87.4494$ was fixed by the mass of the deuteron.
 \subsection{GFFs for the deuteron at zero temperature} 
	
	Following Refs. \cite{4,5,6} we describe deuteron by the twist $\tau=6$ vector field $d_N$ with $d_5=0$. Similarly to the spin-1 particles  the matrix elements of the light-front energy-momentum stress tensor operator $T^{++}(0)$ between the deuteron's initial and final states are expressed by the gravitational form factor $\mathcal{T}^+_{\lambda_2\lambda_1}(Q^2) $\cite{3,8}:
	\begin{eqnarray} \label{3}
	&&\left<p^+, \frac{\bf{q}_\perp}{2},\lambda_2\Big| T^{++}(0) 
	\Big|p^+,-\frac{\bf{q}_\perp}{2},\lambda_1\right>\nonumber\\
	&\!=\!&
	2(p^+)^2  \,  e^{i(\lambda_1-\lambda_2)\phi_q}
	\mathcal{T}^+_{\lambda_2\lambda_1}(Q^2).
	\end{eqnarray}	
	
	Fourier transform of the $\mathcal{T}^+_{\lambda_2\lambda_1}(Q^2)$ form factor gives the $p^+$  gravitomagnetic momentum density  in the transverse plane ($p^+=\frac{p^0+p^3}{\sqrt{2}}$). Here  $\lambda_1$ and $\lambda_2$ are the light-front helicities of the deuteron's initial and final states, ${\bf{q}_\perp}=Q(\cos \phi_q\hat{e}_x+\sin \phi_q\hat{e}_y)$. We shall use the unpolarized case, where $\Big|{\bf{q}_\perp}\Big|=Q$ for the $\lambda_{1,2}=1$ and $\lambda_{1,2}=0$ states.
	
There are two independent helicity conserving FFs $\mathcal{T}^+_{00}$ and $\mathcal{T}^+_{11}(Q^2)$, which are equal to the $Z_1$ and $Z_2$ functions respectively, and have the forms below \cite{3,8}:
\begin{eqnarray}
\mathcal{T}^+_{00}(Q^2)=Z_1(Q)&=&\int \frac{dz}{z} e^{-k^2z^2} H(Q,z)\partial_z\psi_n(z) \partial_z\psi_n(z) \,,
            \nonumber \\
\mathcal{T}^+_{11}(Q^2)=Z_2(Q)&=&\int \frac{dz}{z} e^{-k^2z^2} H(Q,z) \psi_n(z) \psi_n(z) \,.    \label{6}
\end{eqnarray}

	Here $H(Q,z)$ is the bulk-to-boundary propagator for the graviton, and $\psi_n(z)$ is the $n$-th mode's profile function in the deuteron $d_N$ wave function's Kaluza-Klein decomposition. An explicit expression for the $H(Q,z)$ propagator was obtained in Ref. ~\cite{1}:
	\begin{equation}\label{7}
	H(Q,z)=\Gamma(a+2) U(a,-1;2k^2z^2)
		=a(a+1) \int_0^1 dx \, x^{a-1} (1-x) \exp\left(-\frac{2k^2z^2 x}{1-x}\right).
	\end{equation}
 In Ref.~\cite{24} it was reduced to the following form:
 \begin{equation}
	H(Q,z)=4\int_{0}^1 dx k^4 z^4\frac{x^{a+1}}{(1-x)^3}\exp\left(-\frac{2k^2z^2x}{1-x}\right),\label{72} 
	\end{equation}
	where $a=\frac{Q^2}{8k^2}$. The $\psi_n(z)$ profiles are found by solving the equation of motion for the deuteron field and have been expressed in terms of generalized Laguerre polynomials $L^{(\alpha)}_n$ in Refs. \cite{4,5}: 
	\begin{equation} \label{8}
	\psi_n(z) = \sqrt{\frac{2\Gamma(n+1)}{\Gamma(n+L+5)}}k^{L+5} \, z^{L+9/2} L_n^{L+4}(k^2z^2).     
	\end{equation}
	We shall consider the ground state of the deuteron, i.e. $n=0$, $L=0$, and $L_{0}^{4}=1$. Using these expressions for $\psi(z)$ and $H(Q,z)$  the integral over $z$ and $x$ in the $Z_{1,2}(Q)$ form factors can be taken and their explicit expressions are found in the following:

\begin{eqnarray}
   && Z_1(Q)=\Big\lbrace{256 k^4 (16 k^2 + Q^2) (24 k^2 +Q^2) (32 k^2 + Q^2) (40 k^2 + Q^2)}\Big \rbrace ^{-1}  \Big\lbrace(40 k^2 + Q^2) (3342336 k^{12} \nonumber\\
   &&+ 81920 k^{10} Q^2 + 159104 k^8 Q^4+15600 k^6 Q^6 + 1284 k^4 Q^8 + 40 k^2 Q^{10} + Q^{12})+2 Q^2 (8 k^2 + Q^2) (18432 k^8 \nonumber\\
   &&+ 3584 k^6 Q^2 + 512 k^4 Q^4 +  16 k^2 Q^6  + Q^8) {}_{2}F_1 (1, 2 + a,  7 + a, -1) \Big \rbrace, 
\end{eqnarray}
\begin{eqnarray}
   && Z_2(Q)=\Big \lbrace {4 k^2 (16 k^2 + Q^2) (24 k^2 + 
     Q^2) (32 k^2 + Q^2) (40 k^2 + Q^2) (48 k^2 + Q^2)} \Big \rbrace ^{-1} \Big \lbrace (48 k^2 + Q^2) (1966080 k^{10} \nonumber\\
&&- 177152 k^8 Q^2 + 8576 k^6 Q^4 - 672 k^4 Q^6 + 4 k^2 Q^8 - Q^{10})+2 Q^2 (8 k^2 + Q^2) (18432 k^8 \nonumber\\
 &&+ 3584 k^6 Q^2 + 512 k^4 Q^4 +   16 k^2 Q^6  + Q^8) {}_{2}F_1 (1, 2 + a,  7 + a, -1) \Big \rbrace. 
\end{eqnarray}
 
	The gravitational root mean square (RMS) radius is defined taking the derivative of the GFF $Z_2(Q^2)$ \cite{1}:
	\begin{equation} \label{9}
	\left\langle r^2 \right\rangle_{\rm grav}= -6 \left. \frac{\partial Z_2(Q)}{\partial Q^2} \right|_{Q^2=0} . 
	\end{equation}
	Value of the gravitational RMS radius is found from $Z_2$ by use of this definition:
	\begin{equation} \label{10}
	\left\langle r^2 \right\rangle_{\rm grav}=1.605~ {\rm fm^2}. 
	\end{equation}
    For the radius, we find $r_{\rm grav}=\sqrt{\left\langle r^2 \right\rangle_{\rm grav}}=1.26 ~{\rm fm}$.  In CLAS experiment of electroproduction of $\phi$ meson in the process $\gamma+d\rightarrow d+ \phi$  in two different ranges of photon energies $E_{\gamma}=1.6- 2.6~{\rm GeV}$ and $E_{\gamma}=2.6- 3.6~{\rm GeV}$ it were found values for the deuteron mass radius $r_{exp}=1.94 \pm 0.45 ~{\rm fm}$ and $1.78 \pm 0.38 ~ {\rm fm}$ respectively, \cite{44}. $r_{\rm grav}$ found here close to the lowest  experimental value for the mass radius $r_{exp}$. 
 
	\subsection{Deuteron GFFs at finite temperature}\label{PDFs}
	
	In Refs. \cite{19,20} the soft-wall model was extended to the finite temperature case by introducing the warping factor $\exp[-\phi(z, T)]$, which depends on temperature. The temperature dependence in such a model was achieved by considering the dilaton constant $k$ as the temperature-dependent one, i.e. the dilaton field in this model is $\phi(z, T)=K^2(T)z^2$. The action for the soft-wall model is written in terms of such a dilaton field:
	\begin{equation}  
	S=\int_{0}\limits ^{\infty} d^4x dz\sqrt{g} e^{-\phi(z,T)}\mathcal{L}\left(x,z,T\right).
	\label{11}
	\end{equation} 
	Here $\mathcal{L}\left(x,z, T\right)$ is the temperature-dependent Lagrangian, and $g=|\det g_{MN}|$ is the determinant of the $g_{MN}$ metric of the AdS-Schwarzschild space with line element:
	\begin{equation} 
	ds^{2}=e^{2A(z)}\left[-f(z)dt^{2}-(d\vec{x})^2-\frac{dz^{2}}{f(z)}\right], \quad
	f(z)=1-\frac{z^{4}}{z_H^{4}}.
	\label{12}
	\end{equation}
	 Here $A(z)=-\log z$, $z_H$ is the position of the event horizon, which defines the temperature $T=\frac{1}{\pi z_H}$.
	In this model it is suitable to apply the Regge-Wheeler tortoise coordinate $r$ related to $z$ and the thermal factor $f_T(r)$ in terms of the $r$ coordinate has the same form as in $z$ coordinate, \ Ref. \cite {19}, i.e.  
	\begin{equation}
	f_T(r)=1-\frac{r^{4}}{z_H^{4}}.
	\label{13}
	\end{equation}
	In terms of the $r$ coordinate the metric of the AdS-Schwarzschild space-time is written with $A(r)=-\log r$ as below:
	\begin{equation} 
	ds^{2}=e^{2A(r)}f^{\frac{3}{5}}_T(r) \left[dt^{2}-\frac{(d\vec{x})^{2}}{f_T(r)}-dr^{2}\right]
	\label{14}
	\end{equation} 
	 and the dilaton field $\phi(r,T)$ becomes: 
	\begin{eqnarray} 
	\phi(r,T)=K^2(T)r^2, \quad K^2(T)=k^2[1+\rho(T)], \\
 \rho(T)=\delta_{T1}\frac{T^2}{12F^2}+\delta_{T2}\left(\frac{T^2}{12F^2}\right)^2,	
	\label{15}
	\end{eqnarray}
	where  $\delta_{T1}=-\frac{N_f^2-1}{N_f}$ and $\delta_{T2}=-\frac{N_f^2-1}{2N_f^2}$ are the constants, and $F=k\frac{\sqrt{3}}{8}$  is the pseudoscalar meson decay constant at zero temperature and in the chiral limit, and $N_f$ is the number of quark flavors \cite{19,20}.
	
	In the case of zero temperature, the deuteron profile function was studied within the soft-wall model in \cite{4,5}. Following the references \cite{19,20}, where the meson and baryon profile functions were derived at finite temperature, we derive the deuteron profile function in the finite temperature case within the soft-wall model. Similarly to the $T=0$ case, we write action for the bulk vector field $d$ with the twist $\tau=6$ and spin $J=1$, which describes the deuteron field at the UV boundary of the  AdS-Schwarzchild space-time: 
	\begin{eqnarray} \label{16}
	S=\int\limits _{0}^{\infty}d^4 x dr e^{-\phi(r,T)} \sqrt{g}\left[-(D^M d_N(x,r,T))^{+}D_{M} d^{N}(x,r,T)+ \nonumber \right.\\
	+\left(\mu^{2}(r,T)+V(r,T)\right) d_M^+(x,r,T)d^{M}(x,r,T) \left. \right].
	\end{eqnarray}
	The covariant derivative $D^{M}=\partial^{M}-ieV^{M}(r,T)$ in our free deuteron field case is just $\partial^M$. The thermal dilaton potential $V(r,T)$ for the $d^M$ field has a form:
	\begin{equation} \label{17}
	V(r,T)=\frac{4}{f_T^{3/5}(r)}K_{T}^2r^2,
    \end{equation}
	and the temperature-dependent bulk mass $\mu^{2}(r,T)$ is related to the zero-temperature mass $\mu^2$:
 \begin{equation} 
 \mu^{2}(r,T)=\frac{\mu^{2}}{f^{3/5}_T(r)}.\label{18}
 \end{equation}
 
	We use the axial gauge condition for the $d$ field $d^{z}\left(x,r,T\right)=0$. Notice that $\partial_{\alpha} \partial^{\beta} d^{\nu}\left(x,r,T\right) \eta^{\alpha\beta}=-M^{2} d^{\nu}\left(x,r,T\right)$, where $M$ is the mass of the deuteron.
	Kaluza-Klein (KK) decomposition for the $d^{\nu}$ field is written as
	\begin{equation} \label{19}
	d^{\nu}\left(x,r,T\right)=e^{(\phi(r,T)- A\left(r\right))/2}\sum_{n}d_{n}^{\nu}\left(x\right) \psi_{n}\left(r,T\right)
	\end{equation}
	and the $d_{n}^{\nu}\left(x\right)$ tower of KK components dual to the deuteron states with the radial quantum number $n$.
	 
	The EOM for the $\psi_{n}\left(r,T\right)$ mode will be obtained from the action (\ref{16}). Explicitly, the equation of motion for the deuteron modes can be written from the general equation for meson modes with spin $J$, taking $J=1$ in it \cite{4}. This equation is the Schr\"{o}dinger-type equation:
	\begin{equation}\label{20}
	\left[-\frac{d^{2}}{dr^{2}}+k^4r^2\left(1+2\rho(T)\right)+\frac{4m^2-1}{4r^2}-4k^2\left(1+\rho(T)\right)\left(n+\frac{m+1}{2}\right)\right]\psi_{nm}\left(r,T\right)=0.
	\end{equation}
	Here $m=4+L$. 
	Solution to this equation - profile function $\psi_{nm}\left(r, T\right)$ for the $n$-th mode of the deuteron has a form:
    \begin{equation}\nonumber
    \psi_{nm}(r,T)=\sqrt{\frac{2\Gamma(n+1)}{\Gamma(n+m+1)}}K_T^{m+1}r^{m+1/2}e^{-K_T^2r^2/2}L_n^m\left(K_T^2r^2/2\right).
    \end{equation}
    For the ground state with $n=0$ and $L=0$, we have the finite-temperature profile function below:
	\begin{equation} \label{21}
	\psi(r,T) = \sqrt{\frac{2}{4!}}K_T^{5} \, r^{9/2}  e^{-K_T^2 r^2/2}.    
	\end{equation} 
	This solution coincides with one obtained in Ref. \cite{8} for the zero temperature case, by replacements $K_T\rightarrow k$ and $r \rightarrow z$ in (\ref{21}). Analogously to zero-temperature case, the form-factors $Z_1(Q^2, T)$ and $Z_2(Q^2, T)$ are written in terms of the deuteron profile function $\psi(r, T)$ and graviton bulk-to-boundary propagator $H(Q, T,r)$: 
	\begin{eqnarray}
	\mathcal{T}^+_{00}(Q^2,T)=Z_1(Q^2,T)&=&\int dr H(Q,T,r)\partial_r\psi(r,T) \partial_r\psi(r,T) \,,
	\nonumber \\
	\mathcal{T}^+_{11}(Q^2,T)=Z_2(Q^2,T)&=&\int dr H(Q,T,r)\psi(r,T) \psi(r,T) \,.    \label{22}
	\end{eqnarray}
	
	The finite-temperature graviton bulk-to-boundary propagator $H(Q, T,r)$ was obtained by solving the EOM at finite temperature, which is obtained from the EOM at zero temperature by the replacements $k \rightarrow K_{T}$ and $z \rightarrow r$. So, the solution $H(Q, T,r)$ differs from the $H(Q,z)$ by this replacement (\cite{24}):
	\begin{eqnarray} \label{23}
	H(Q,T,r)=\Gamma(a_{T}+2) U(a_{T},-1;2K_T^2r^2)= 
	\nonumber \\
	=a_{T}(a_{T}+1) \int_0^1 dx \, x^{a_{T}-1} (1-x) \exp\left(\frac{-2K_T^2r^2 x}{1-x}\right)= \nonumber\\
 =4\int_{0}^1 dx K_T^4 r^4\frac{x^{a_T+1}}{(1-x)^3}\exp\left(\frac{-2K_T^2r^2x}{1-x}\right),
	\end{eqnarray}
	where $a_T=\frac{Q^2}{8K_T^2}$.
	Having plotted the $Q^2$ and $T$ dependencies of the $Z_{1,2}(Q^2, T)$ form factors in Fig. 1 and Fig. 2 we visualize the temperature impact on the energy-momentum distribution of partons inside the deuteron.    
	\begin{figure*}[htbp]
		\begin{minipage}[c]{0.98\textwidth}
			{(a)}\includegraphics[width=7.5cm,clip]{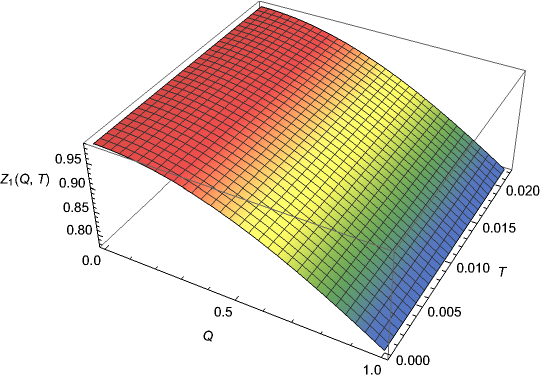}
			{(b)}\includegraphics[width=7.5cm,clip]{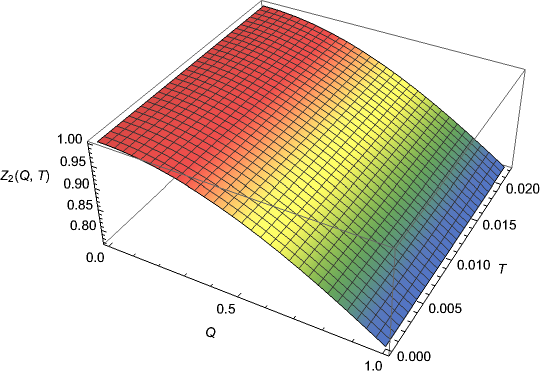}
		\end{minipage}
		\caption{ Plots of (a) $Z_{1}(Q^2,T)$ form factor and (b) $Z_{2}(Q^2,T)$ form factor at finite temperature for the deuteron in $\rm fm$ units.}
		\label{}
	\end{figure*}
	The plots for the GFFs in Fig.1 show very weak temperature dependences of these form factors.  This means that temperature weakly affects the distribution of energy and momentum inside the deuteron.

Using the expression of finite temperature for $Z_{2}(Q^2, T)$ we can find the gravitational radius of the deuteron at finite temperature by taking the derivative of this gravitational form factor:
	\begin{equation} \label{24}
	\left\langle r^2(T) \right\rangle_{\rm grav}= -6 \left. \frac{\partial Z_2(Q,T)}{\partial Q^2} \right|_{Q^2=0} . 
	\end{equation}
	\begin{figure}[htbp]
  \includegraphics[width = 0.4\textwidth]{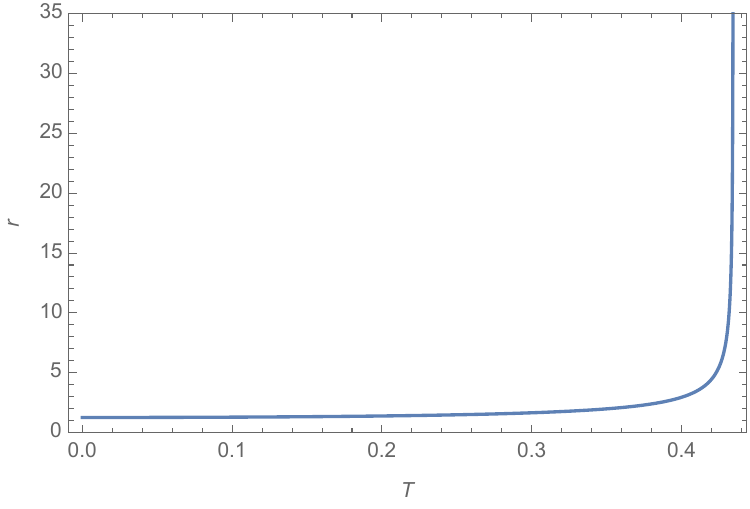}
		\caption{Plot of gravitational RMS radius for the deuteron.}
		\label{}
	\end{figure}
	
	As is seen from Fig. 2 the RMS radius plot increases on increasing medium temperature starting from $r_{\rm grav}=1.26$ $\rm fm$ at $T=0$ value. Starting from the temperature $T\approx 0.21~{\rm fm}^{-1}$ the radius grows quickly and becomes infinite at the temperature $T\sim 0.42~{\rm fm}^{-1}$. This graph behavior can be related to the change in structure in the deuteron. As is known \cite{book}, the binding energy for the deuteron is $\sim T=2.22~{\rm MeV}$ $\simeq 0.11~{\rm fm}^{-1}$ and it does not have excited states. At a temperature equal to $T\simeq 0.11~{\rm fm}^{-1}$, the deuteron splits into proton and neutron and does not behave as a single particle. So, only $T<0.11~{\rm fm}^{-1}$ part of this graph describes the deuteron.  
	
	\bigskip
	\section{Deuteron GPDs}
		
	GPDs describe the distributions of partons inside the hadron and are functions of the longitudinal momentum fraction $x$ of the active quark or gluon, skewness $\xi$ and $Q^2$. Here we shall not study the skewness dependence of the GPDs. Within the hard-wall model of AdS/QCD the GPDs for the vector field were written in Ref. \cite{1} and were related to holographic GFFs. (The study of deuteron's the GPDs within the soft-wall model was done in Ref. \cite{8} with the EM FFs study.) We aim to obtain the deuteron of the GPDs from the GFFs using the relation between them, which was applied in Ref.~\cite{1} for the vector field. The GFFs study for the vector field was extended to the soft-wall model and the non-zero temperature case in Ref. \cite {24}. Similar research can also be done for the deuteron, taking the twist of the vector field $\tau=6$. For the $n$-th deuteron state $d_{n}$ the GFFs are defined by the following matrix element of $\hat{T}^{\mu\nu}$ - the transverse-traceless part of the energy-momentum stress tensor $T^{\mu\nu}$:
	
	\begin{eqnarray}
\left<d_n^a(p_2,\lambda_2)\big|\hat{T}^{\mu\nu}(q)\big|d_n^b(p_1,\lambda_1)\right>=\nonumber \\
\quad (2\pi)^4 \delta^{(4)}(q+p_1-p_2) \, \delta^{ab} \,
	\varepsilon^*_{2 \alpha} \varepsilon_{1 \beta} \nonumber\\[1ex]
\times\bigg[ - A(q^2)\bigg(4 q^{[\alpha} \eta^{\beta](\mu} p^{\nu)}
	+2\eta^{\alpha\beta}p^\mu p^\nu\bigg)				\nonumber\\
\quad-\ \frac{1}{2} \hat C(q^2)
\eta^{\alpha\beta}\bigg(q^2 \eta^{\mu\nu} - q^\mu q^\nu  \bigg) \nonumber\\
\quad +\  D(q^2) \bigg( q^2 \eta^{\alpha(\mu}\eta^{\nu)\beta}
	- 2q^{(\mu}\eta^{\nu)(\alpha}q^{\beta)}
		+ \eta^{\mu\nu} q^\alpha q^\beta \bigg) \nonumber\\
\quad-\ \hat F(q^2) \frac{q^\alpha q^\beta}{m_n^2}
	\bigg( q^2 \eta^{\mu\nu}- q^\mu q^\nu \bigg)\bigg] \,, 
				\label{4}
\end{eqnarray}
where $m_{n}$ is the mass of the $d_n$ state. The FFs $A, \hat C, D, \hat F$ are composed from the invariant functions $Z_1$ and $Z_2$:
\begin{eqnarray}
A(q^2)&=&Z_2   \,,  \nonumber \\
\hat C(q^2)&=&  \frac{1}{q^2}
    \bigg(\frac{4}{3}Z_1+\big( q^2 - \frac{8m_n^2}{3}\big)Z_2\bigg), \nonumber \\
D(q^2)&=& \frac{2}{q^2} Z_1+\left( 1- \frac{2m_n^2}{q^2} \right)Z_2, \nonumber \\
\hat F(q^2)&=& \frac{4m_\rho^2}{3q^4}  \left( Z_1 - m_n^2 Z_2 \right).    \label{5}
\end{eqnarray}

	Generally, there are five the GPDs for the spin-1 particles, which are expressed with the matrix element of the QCD non-local and non-diagonal operators
	on the light-front \cite{1,8,42}: 
	\begin{eqnarray}
		&& \int \frac{p^+ dy^-}{2\pi} e^{ixp^+y^-} \times 
		\nonumber \\ 
		&& \quad	\left\langle p_2,\lambda_2 \right|  \bar q(-\frac{y}{2}) \gamma^+  
		q(\frac{y}{2})\left| p_1, \lambda_1 \right\rangle_{y^+=0, y_\perp = 0}
		\nonumber \\
		&& = - 2 (\varepsilon_2^* \cdot \varepsilon_1) p^+ H_1(x,\xi,t)
		- \left( \varepsilon_1^+ \, \varepsilon_2^* \cdot q 
		- {\varepsilon_2^+}^* \, \varepsilon_1 \cdot q \right) H_2(x,\xi,t)
		\nonumber \\
		&& \quad + \ q\cdot \varepsilon_1 \, q\cdot \varepsilon_2^* \frac{p^+}{m_n^2} H_3(x,\xi,t)
		- \left( \varepsilon_1^+ \, \varepsilon_2^* \cdot q 
		+ {\varepsilon_2^+}^* \, \varepsilon_1 \cdot q \right) H_4(x,\xi,t)
		\nonumber \\
		&& \quad +\ \left( \frac{m_n^2}{(p^+)^2} \varepsilon_1^+ \, {\varepsilon_2^+}^*
		+ \frac{1}{3} (\varepsilon_2^* \cdot \varepsilon_1) \right) 2 p^+ \, H_5(x,\xi,t) \ , \label{25}
	\end{eqnarray}
	where $x$ is a momentum fraction $x=\frac{k^+}{p^+}$ 
	varying in the interval $-1\leq x \leq 1$, $k^+$ is the parton’s (quark’s) light-cone momentum, $\xi$ is a skewness $\xi=-\frac{q^+}{2p^+}$, and valence quark the GPDs $H_i^v$ are defined by $H_i$ as $H_i^v(x,0,t)=H_i(x,0,t)+H_i(-x,0,t)$.
	
	The first moments of deuteron's $H_i$ the GPDs for the valence quark are related to the linear combinations of the $A, B, C, D, E$ FFs in (\ref{4}) in the following \cite{1,42}:
\begin{align}
\int_{-1}^1 x dx \, H_1(x,\xi,t) &=A(t) - \xi^2 C(t) + \frac{t}{6m_\rho^2} D(t)	,	\nonumber \\
\int_{-1}^1 x dx \, H_2(x,\xi,t) &= 2 \left( A(t)+B(t) \right)		,	\nonumber \\
\int_{-1}^1 x dx \, H_3(x,\xi,t) &= E(t) + 4 \xi^2 F(t)				,	\nonumber \\
\int_{-1}^1 x dx \, H_4(x,\xi,t) &= -2 \xi D(t)					,	\nonumber \\
\int_{-1}^1 x dx \, H_5(x,\xi,t) &=  + \frac{t}{2m_\rho^2} D(t)			\label{26}	.	
\end{align}

Using the explicit expression for $\psi(z)$ in Eq.~(\ref{6}) and $H(Q, z)$ in Eq.~(\ref{72}) the integral expression for the $Z_1, Z_2$ the GFFs can be found:
	\begin{eqnarray}
		Z_1(Q)&=&\int_0^1 dx \int_0^\infty dz  k^{14} z^{11} \frac{8x^{(a+1)}}{(1-x)^3} e^{\frac{-k^2z^2(1+x)}{1-x}}    \,,
            \nonumber \\
		Z_2(Q)&=&\int_0^1 dx \int_0^\infty dz  k^{14} z^{13} \frac{x^{(a+1)}}{3(1-x)^3} e^{\frac{-k^2z^2(1+x)}{1-x}}.    \label{27}
	\end{eqnarray}
	 Taking into account Eq. (\ref{27}) in Eq.~(\ref{5}) and the obtained result in Eq. (\ref{26}) we obtain an expression  for the deuteron the GPDs for $\xi=0$:
	\begin{eqnarray}\label{28}
		H_1^v(x,0,t)=-\frac{5(1-x)^3 (4(4m^2+t)(1-x)+27k^2(1+x))}{m^2(1+x)^7} x^{-a},\
            \nonumber \\
		H_2^v(x,0,t)=-\frac{240(x-1)^4}{(1+x)^7} x^{-a}, \,
            \nonumber \\
H_5^v(x,0,t)=-\frac{15(x-1)^3 (4(2m^2-t)(x-1)+27k^2(1+x))}{m^2(1+x)^7} x^{-a},
	\end{eqnarray}
	where $t=-Q^2$.

	\begin{figure*}[htbp]
		\begin{minipage}[c]{0.98\textwidth}
			{(a)}\includegraphics[width=7.5cm,clip]{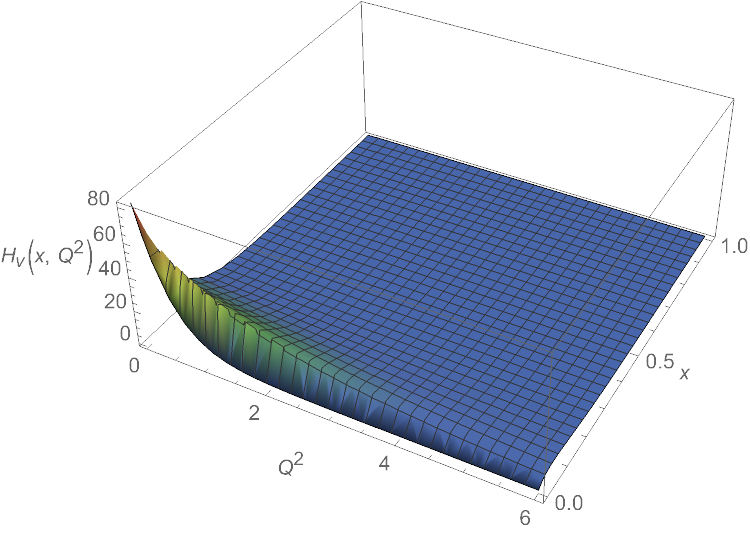}
			{(b)}\includegraphics[width=7.5cm,clip]{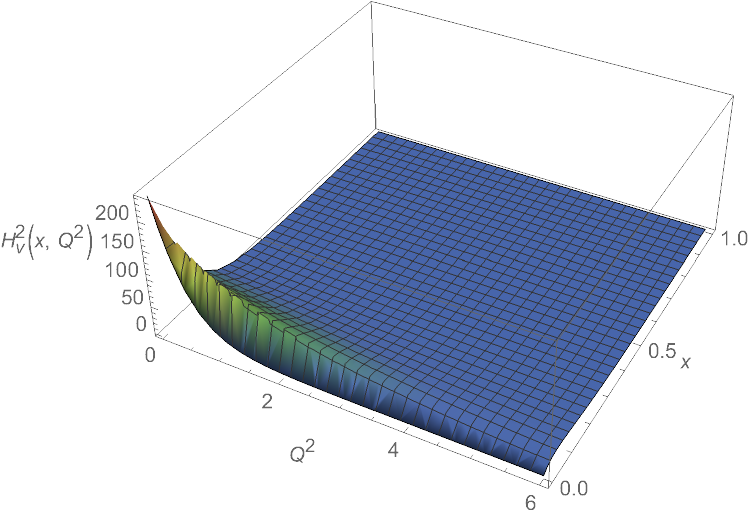}
			{(c)}\includegraphics[width=7.5cm,clip]{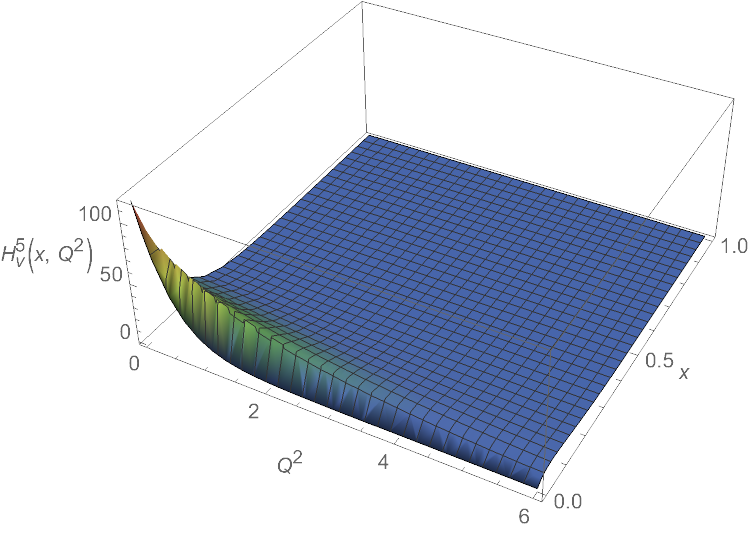}
            {(d)}\includegraphics[width=7.5cm,clip]{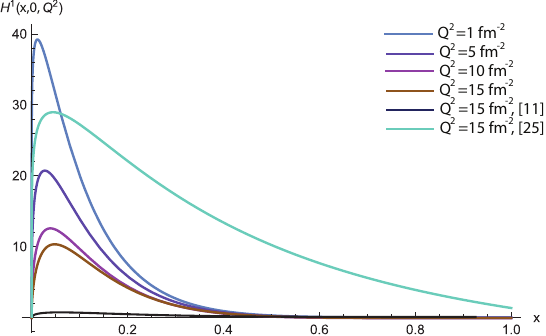}
		\end{minipage}
		\caption{Plot of $H_v^i(x,0,t)$  is generalized parton distributions for deuteron.}
		\label{H(xQ)}
	\end{figure*}
It is seen from Fig. 3 that the $H^i_v$ GPDs have a peak near the $(x=0, Q^2=0)$ point. Such a peak and shape have the GPDs obtained from the EM FFs of the deuteron in Ref. \cite{8}.

	The deuteron $H_i$ GPDs at zero temperature can easily be extended to the finite-temperature case by applying the temperature-dependent propagator $H(Q, T,r)$. An integral representation for $H(Q, T,r)$ is obtained by replacing the dilaton constant $\kappa$ by $K_T$ and the $z$ coordinate by the $r$ tortoise one in the zero-temperature expression (\ref{72}):
	\begin{eqnarray} \label{29}
	H(Q,T,r)=4\int_{0}^1 dx K_T^4 r^4\frac{x^{a_{T}+1}}{(1-x)^3}e^{\frac{-2K_T^2r^2x}{1-x}}. 
	\end{eqnarray}	
	So, finite-temperature version of the GPDs (\ref{27}) will be written as:
	\begin{eqnarray} \label{30}
	H_1^v(x,0,t,T)=-\frac{5(1-x)^3 (4(4m^2+t)(1-x)+27K_T^^2(1+x))}{m^2(1+x)^7} x^{-a_T}. 
	\end{eqnarray}
 The GFFs in terms of $H(Q,T,r)$ read:
	\begin{eqnarray}
		Z_1(Q,T)&=&\int_0^1 dx \int_0^\infty dr   K_T^{14} r^{11} \frac{8x^{(a_T+1)}}{(1-x)^3} e^{\frac{-K_T^2r^2(1+x)}{1-x}}   \,,
            \nonumber \\
		Z_2(Q,T)&=&\int_0^1 dx \int_0^\infty dr  K_T^{14} r^{13} \frac{x^{(a_T+1)}}{3(1-x)^3} e^{\frac{-K_T^2r^2(1+x)}{1-x}}.    \label{31}
	\end{eqnarray}

	\begin{figure*}[htbp]
		\begin{minipage}[c]{0.98\textwidth}
			{(a)}\includegraphics[width=7cm,clip]{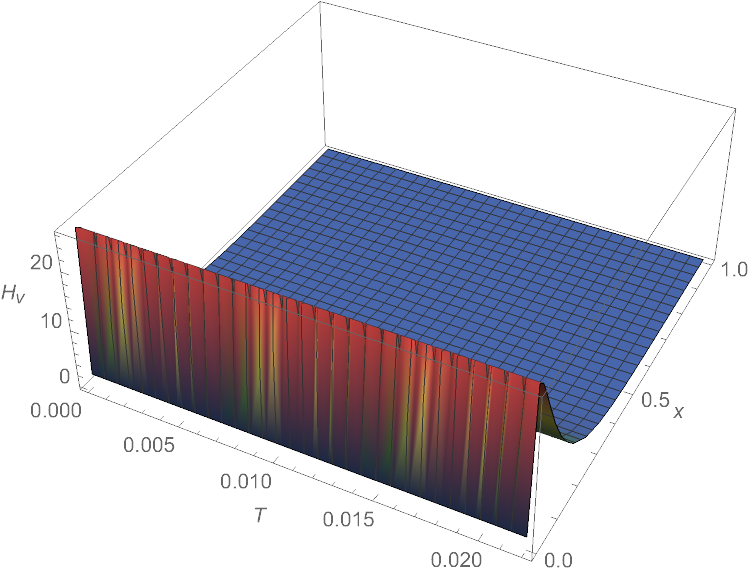}
			{(b)}\includegraphics[width=7cm,clip]{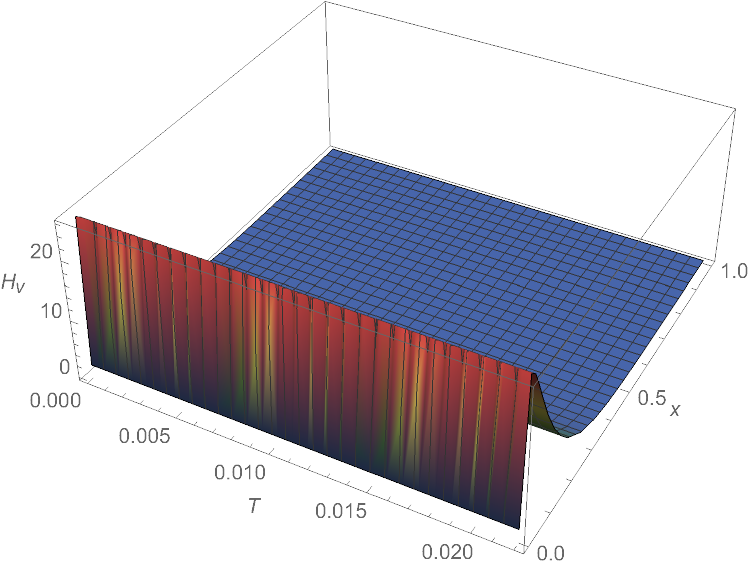}
		\end{minipage}
		\caption{Plot of $H_v^1(x,0,t,T)$  is generalized parton distributions at finite temperature. Accordingly, (a) is $Q^2=1~{\rm fm^{-2}}$ and (b) is $Q^2=3~{\rm fm^{-2}}$.}
		\label{}
	\end{figure*}
	
	It is seen from Fig.4, both GPDs have a maximum at $x\approx 0.2$ and temperature weakly affects these maxima. Such a maximum can be met in Ref. \cite{45} (Fig. 1), where the GPDs were studied for the $u$ and $d$ quarks at zero temperature within the hard-wall model. A holographic study of the deuteron GPDs obtained from the EM FFs within the soft-wall model indicates the presence of the deuteron GPDs maxima near $x\approx 0.2$ (\cite{8}, Fig. 8 (b)). In addition, the plot of vector meson GPDs has such a maximum (\cite{24}). There is such a peak in zero-temperature plots in Fig. 3 (d), as well.
	\bigskip
 
	\section{Deuteron GPDs in IP space }
		
	The IP-dependent parton distribution functions (PDFs) for a quark of momentum fraction $x$ located in the transverse position $\bf b_\perp$ in the impact parameter space give a picture of the distribution of quark charge densities inside the deuteron through $H_i(x,{\bf b_\perp})$. Using the GPDs in impact parameter space, one can calculate the distributions of partons in the transverse plane, which is significant for the study of deuteron structure. For localizing the deuteron in the transverse direction, one needs to place the transverse center of momentum ${\bf R_\perp}$ at the origin, and such a state in the transverse center of momentum is defined as \cite{43}:
	\begin{eqnarray}
		\left|p^+,{\bf R_\perp}= {\bf 0_\perp}, \lambda\right\rangle \equiv {\cal N}\int \frac{d^2{\bf p_\perp}}{(2\pi)^2}  \left|p^+,{\bf p_\perp}, \lambda \right\rangle. \label{32}
	\end{eqnarray}
	
	The normalization constant ${\cal N}$ is found from the normalization $|{\cal N}|^2\int \frac{d^2{\bf p_\perp}}{(2\pi)^2}=1$. $\left|p^+,{\bf p_\perp}, \lambda \right\rangle$ are light-cone helicity eigenstates of the deuteron. Thus, one can define deuteron GPDs in IP space $H_i(x,{\bf b_\perp})$:
	\begin{eqnarray}
		H(x,{\bf b_\perp}) \equiv 
		\left\langle p^+,{\bf R_\perp}= {\bf 0_\perp},
		\lambda\right|
		\int \frac{dy^-}{4\pi}\bar{q}\left(-\frac{y^-}{2},{\bf b_\perp} \right)  \gamma^{+} q\left(\frac{y^-}{2},{\bf b_\perp}\right)
		\left|p^+,{\bf R_\perp}= {\bf 0_\perp},
		\lambda\right\rangle e^{ixp^+y^-} \nonumber\\
		=|{\cal N}|^2\int \frac{d^2{\bf p_\perp}}{(2\pi)^2}
		\int \frac{d^2{\bf p'_\perp}}{(2\pi)^2}\left\langle p^+,{\bf p'_\perp},
		\lambda\right|\int \frac{dy^-}{4\pi}\bar{q}
		\left(-\frac{y^-}{2},{\bf 0_\perp} \right)
  \gamma^{+}
		q\left(\frac{y^-}{2},{\bf 0_\perp}\right) 
		\left|p^+,{\bf b_\perp},\lambda\right\rangle e^{i{\bf b_\perp}({\bf p_\perp}-{\bf p'_\perp})}e^{ixp^+y^-}. \nonumber\\.
		\label{33}
	\end{eqnarray} 
	
	To illustrate the physics of the GPDs $H_i^v(x,0,q)$, we take the definition of IP-dependent parton distribution functions (PDFs):
	\begin{eqnarray}
		H(x,{\bf b_\perp}) =|{\cal N}|^2\int \frac{d^2{\bf p_\perp}}{(2\pi)^2}
		\int \frac{d^2{\bf p'_\perp}}{(2\pi)^2}H^v(x,0,{\bf q}^2_\perp)e^{i{\bf b_\perp}({\bf p_\perp}-{\bf p'_\perp})} \nonumber \\
		=\int \frac{d^2{\bf q_\perp}}{(2\pi)^2}H^v(x,0,{\bf q}^2_\perp)e^{-i{\bf b_\perp}{\bf q_\perp}},
		\label{34}
	\end{eqnarray}
	where ${\bf b_{\perp}}$ is the IP and $|{\bf b}_\perp|=b$ is a measure of the transverse distance between	the struck parton and the transverse center of momentum of the hadron, $\bf q_\perp=\bf p'_\perp-\bf p_\perp$. As is shown in Ref. \cite{43}, the GPDs in the momentum space are related to the IP-dependent parton distributions via the Fourier transform:
	\begin{eqnarray}
		H_i(x,{\bf b_\perp}) =\int_{0}^{\infty} \frac{d^2{\bf q_\perp}}{(2\pi)^2}H_i^v(x,0,{\bf q}^2_\perp)e^{-i{\bf b_\perp}{\bf q_\perp}}.
		\label{35}
	\end{eqnarray}
After integrating over the angle between the $\bf b_\perp$ and $\bf q_\perp$ vectors $H_i(x,{\bf b_\perp})$ are writing as an integral over $Q=|\bf q_\perp|$:
\begin{equation}
H_{i}\left(x,{\bf b_\perp}\right) =\int_{0}^{\infty} \frac{dQ}{2\pi}Q J_{0} (b Q)H_i^v(x,0,Q^2),
\end{equation}	
where $J_0$ is the cylindrical Bessel function of order zero. After taking into account (\ref{28}) in (\ref{35}) and integrating out over $Q$, we find an explicit expression for the generalized parton distributions $H_i$ in the IP space: 
\begin{align} \label{36}
H_1(x,b)=-\frac{10k^2(x-1)^3 e^{\frac{2b^2k^2}{\lg{x}}}}{\pi m^2(1+x)^7\lg{x}^3} \left(64b^2k^4(x-1)+\lg{x}(32k^2(x-1)+\right. \nonumber\\
\left. (16m^2(x-1)-27k^2(x+1))\lg{x})\right),
  \nonumber \\
H_2(x,b)=-\frac{480k^2(x-1)^4}{(1+x)^7\lg{x}} e^{\frac{2b^2k^2}{\lg{x}}},    \nonumber \\
H_5(x,b)=-\frac{30k^2(x-1)^3}{m^2(1+x)^7\lg{x}^3})e^{\frac{2b^2k^2}{\lg{x}}}(64b^2k^4(x-1)+32k^2(x-1)\lg{x}- \nonumber\\
(8m^2(x-1)+27k^2(1+x)\lg{x}^2).
	\end{align}
A plot of these GPDs can be performed numerically:
	\begin{figure*}[htbp]
		\begin{minipage}[c]{0.98\textwidth}
			\includegraphics[width=7.5cm,clip]{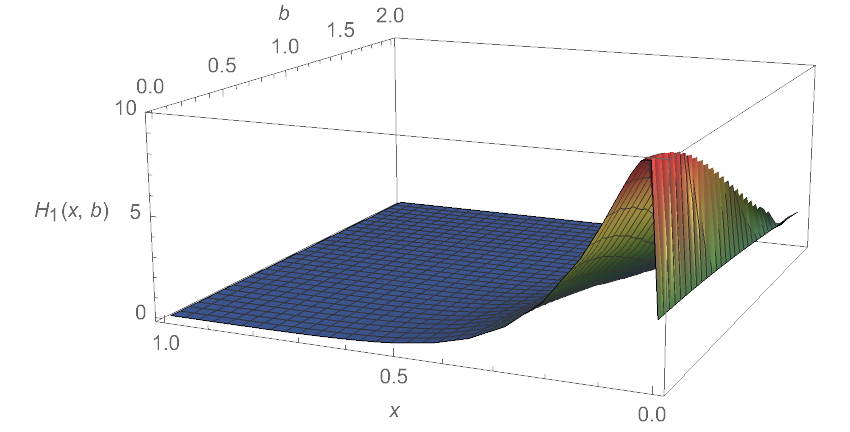}
			\includegraphics[width=7.5cm,clip]{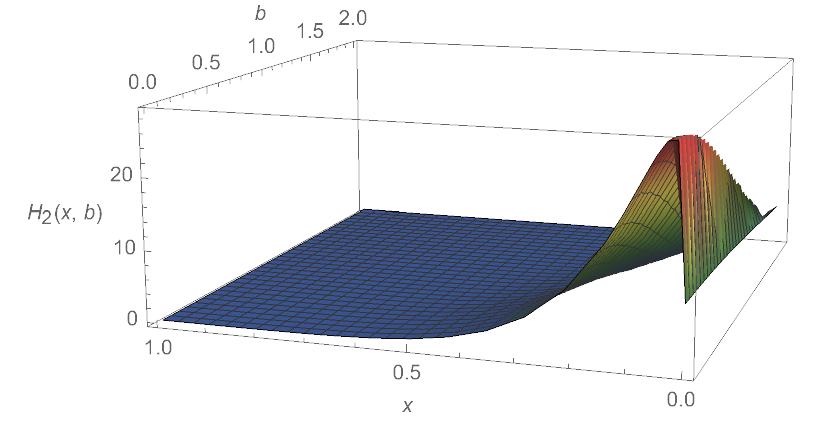}
			\includegraphics[width=7.5cm,clip]{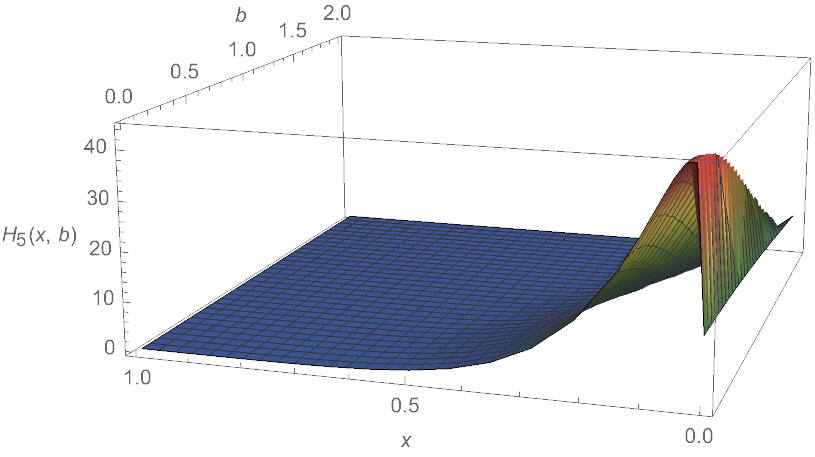}
		\end{minipage}
		\caption{Plot of $H_1(x,b)$, $H_2(x,b)$, $H_5(x,b)$ is generalized parton distributions in impact parameter space.}
		\label{H(x,b)}
	\end{figure*}
	It is seen from the plots in Fig.5 that the peak of the graphs of the GPDs in IP space decreases at an increase of the impact parameter $b$.
	In the thermal dilaton soft-wall model we only replace the dilaton parameter $k$ with the temperature-dependent dilaton parameter $K_T$. This allows us to express the GPDs in the impact parameter space at finite temperature in the same way as in the zero-temperature case. To do this, we replace $k$ in (\ref{36}) by $K_T$:

	\begin{eqnarray} \label{37}
		H_1(x,b,T)=-\frac{10K_T^2(x-1)^3 e^{\frac{2b^2K_T^2}{\lg{x}}}}{\pi m^2(1+x)^7\lg{x}^3} (64b^2K_T^4(x-1)+\lg{x}(32K_T^2(x-1)+ \nonumber\\
(16m^2(x-1)-27K_T^2(x+1))\lg{x})),
            \nonumber \\
		H_2(x,b,T)=-\frac{480K_T^2(x-1)^4}{(1+x)^7\lg{x}} e^{\frac{2b^2K_T^2}{\lg{x}}},    \nonumber \\
H_5(x,b,T)=-\frac{30K_T^2(x-1)^3}{m^2(1+x)^7\lg{x}^3}e^{\frac{2b^2K_T^2}{\lg{x}}}(64b^2K_T^4(x-1)+32K_T^2(x-1)\lg{x}- \nonumber\\
(8m^2(x-1)+27K_T^2(1+x)\lg{x}^2).
	\end{eqnarray}

\begin{figure}[htbp]
			\begin{minipage}[c]{0.98\textwidth}
			\includegraphics[width=7.5cm,clip]{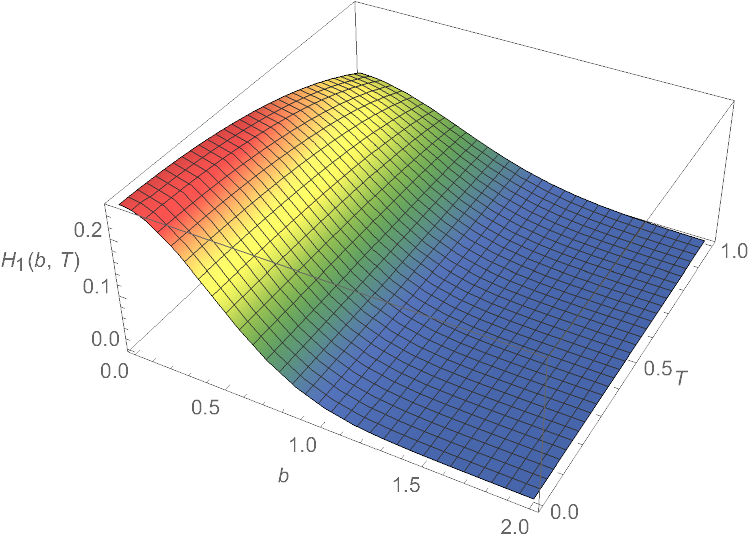}
			\includegraphics[width=7.5cm,clip]{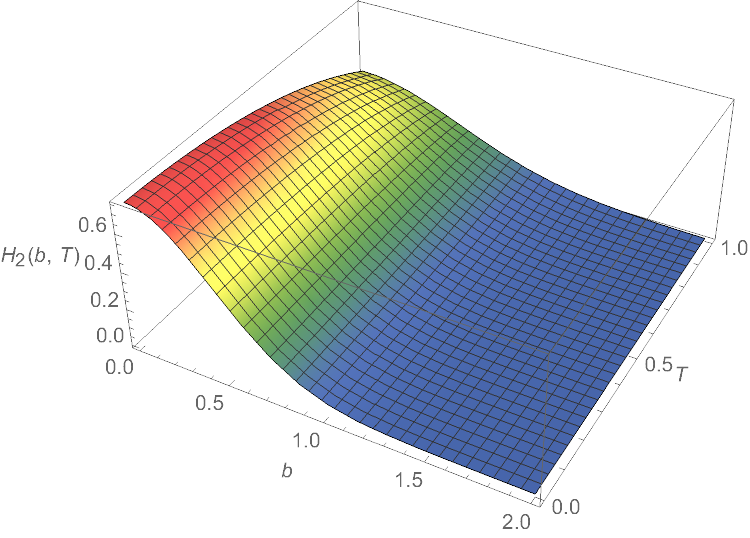}
			\includegraphics[width=7.5cm,clip]{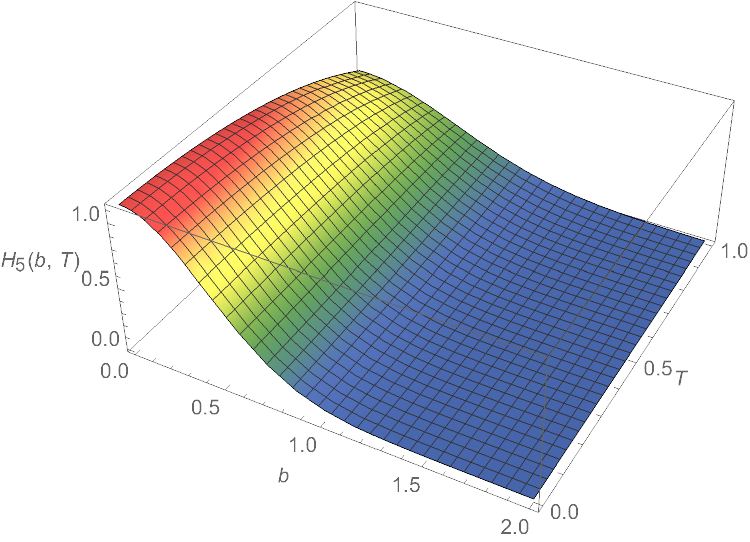}
		\end{minipage}
		\caption{Plot of $H_1(x,b, T)$, $H_2(x,b, T)$, $H_5(x,b, T)$ is generalized parton distributions in impact parameter space at finite temperature in $x=0.5$.}
		\label{}
	\end{figure}
As seen from Fig. 5, the plots of different $H_i$ have similar behavior and peak at $x=0$, despite different expressions in (\ref{37}) for these GPDs. Comparing with the plots in Fig. 4, we observe that GPDs in IP space are more sensitive to temperature, as peaks of graphs in Fig. 5 decrease on temperature rising. 
	\bigskip
	\section{Deuteron transverse charge density}
	 Transverse charge density $\rho^+_\lambda$ in IP space is defined as a Fourier transform of the $\mathcal{T}^+_{\lambda\lambda}(Q^2)$ form factors:
	\begin{equation} \label{38}
		 \rho^+_\lambda({\bf{b}_\perp}) = \int \frac{d^2\bf{q}_\perp}{(2\pi)^2} e^{-i\bf{q}_\perp\cdot\bf{b}_\perp}\mathcal{T}^+_{\lambda\lambda}(Q^2).
	\end{equation}
	It gives a probability that the charge is located at transverse separation $b$ from the transverse center of momentum. The $\rho^+_\lambda$ density of the spin-$1$ particle is related to the IP space 2-dimensional Fourier transform of the $Z_{1,2}$ GFFs \cite{3}. In holographic QCD for the special $\tau=6$ case one can obtain the following charge densities for deuteron \cite{8}:
	\begin{eqnarray} \label{39}
		\rho^+_1(b)&=&\int_0^\infty \frac{dQ}{2\pi}Q J_0(bQ)Z_2(Q^2)\,,\nonumber\\
		\rho^+_0(b)&=&\int_0^\infty \frac{dQ}{2\pi}Q J_0(bQ)Z_1(Q^2).
	\end{eqnarray}
	Using holographic expressions (\ref{27}) one can investigate transverse charge densities (\ref{39}) numerically. In Fig. 7 a, we plot a 3-dimensional graph for the $\rho^+_1$ density, and in Fig.7b, we give its 2-dimensional projection in the ($b_x,b_y$) plane. The shape of these plots coincides with 2-dimensional graphs plotted in the Ref. \cite{8}.
	\begin{figure*}[htbp]
		\begin{minipage}[c]{0.98\textwidth}	 	         
			{(a)}\includegraphics[width=8.5cm,clip]{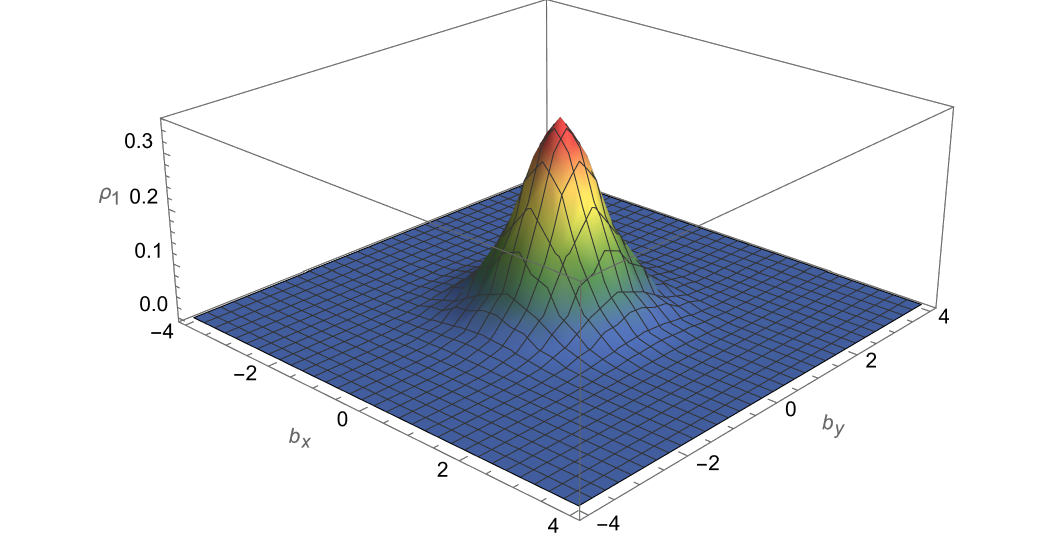}
			{(b)}\includegraphics[width=5cm,clip]{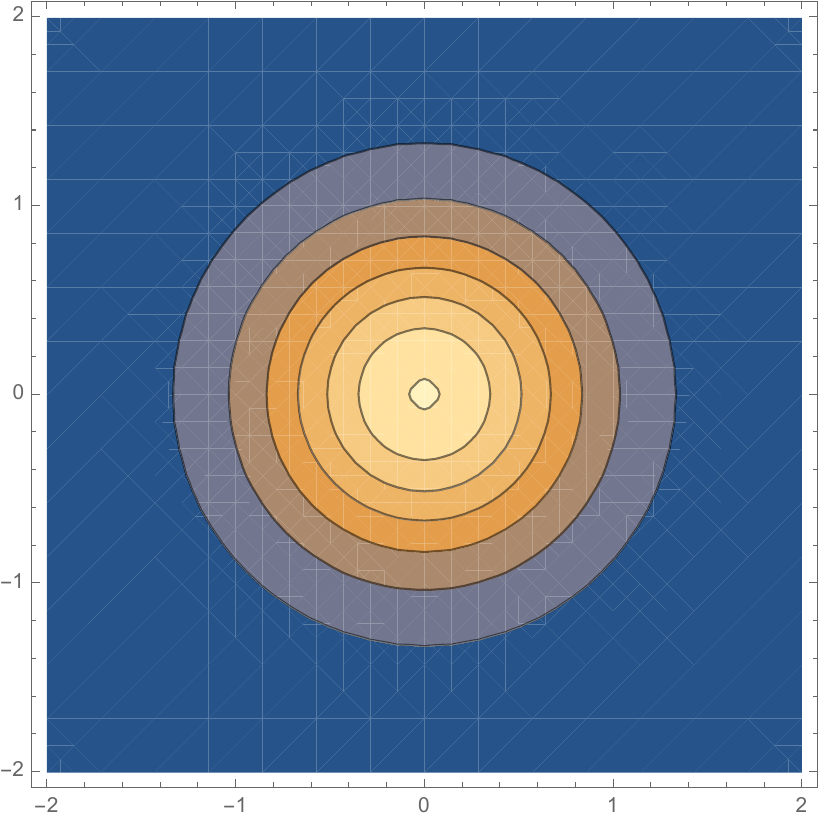}
		\end{minipage}
		\caption{Plot of $\rho^+_1(\vec{b}_\perp)$ transverse charge density of deuteron.}
		\label{H(x,b)}
	\end{figure*}

		In order to have charge density in IP space we have to take into account in (\ref{38}) the finite-temperature $\mathcal{T}^+_{\lambda\lambda}(Q^2)$ in (\ref{22}). This will lead to replacement in (\ref{39}) the $Z_{1,2}(Q^2)$ form factors by the thermal ones $Z_{1,2}(Q^2,T)$:	
		\begin{eqnarray} \label{40}
		\rho^+_1(b,T)&=&\int_0^\infty \frac{dQ}{2\pi}Q J_0(bQ)Z_2(Q^2,T)\,,\nonumber\\
		\rho^+_0(b,T)&=&\int_0^\infty \frac{dQ}{2\pi}Q J_0(bQ)Z_1(Q^2,T).
	\end{eqnarray}
	
\begin{figure}[htbp]
			\begin{minipage}[c]{0.98\textwidth}
			{(a)}\includegraphics[width=7.5cm,clip]{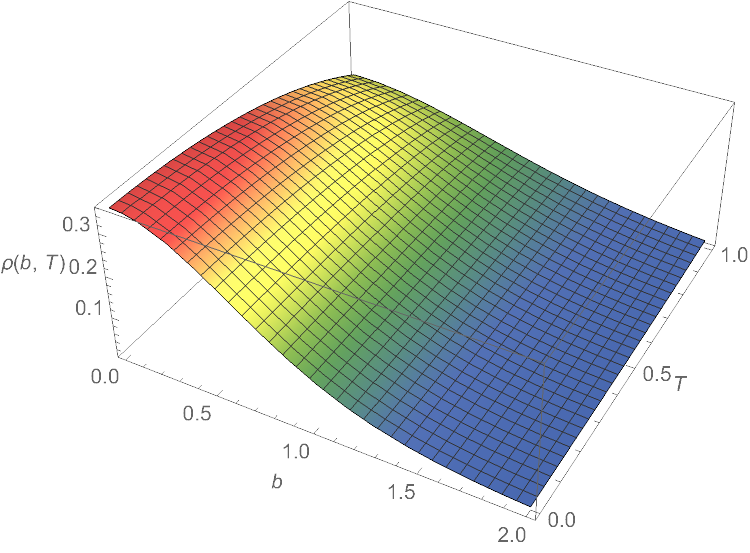}
   {(b)}\includegraphics[width=7.5cm,clip]{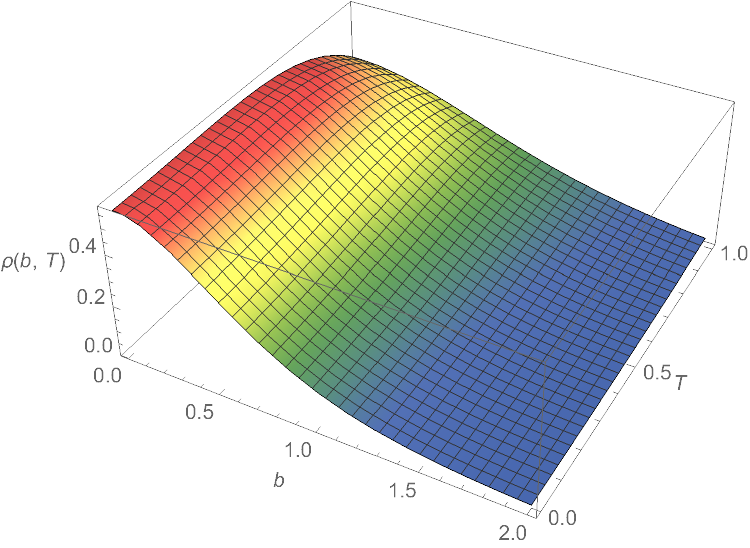}
   {(c)}\includegraphics[width=7.5cm,clip]{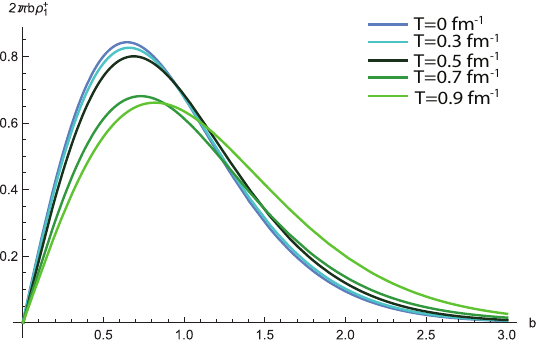}
		\end{minipage}
		\caption{Plot (a) is $\rho^+_1(b,T)$ charge density for $Z_2$ and (b) is $\rho^+_0(b,T)$ charge density for $Z_1$ of deuteron at finite temperature.}
		\label{}
	\end{figure}
As is seen from the graphs in Fig. 8 the charge density peak is higher at the temperature $T=0$ than at $T=1~{\rm fm}^{-1}$. This means that the charge density peak decreases as the temperature rises.
\section{Summary}
The deuteron GPDs found here from the GFFs, have the same shape as the ones obtained from the EM FFs in Ref. \cite{8}. This agreement between two holographic results for the deuteron indicates the usefulness of the holographic approach in the study of the inner structure of hadrons. The RMS radius obtained from the GFFs is close to the experimental data for the mass radius. The temperature dependence study of the radius becomes infinite starting from the critical value of the temperature. The shapes of plots of the GPDs obtained from different form factors are the same. Such shape form is characteristic for the nucleons (\cite{45}), pions (\cite{46}) and vector mesons (\cite{24}). The charge density peak in IP space is weakly sensitive to the temperature change. The study here can be extended to the skewness dependence of the GPDs. The description of the quantities of the deuteron inner structure can also be studied, taking into account the interaction between nucleons. To this end, it is useful to extend the Argonne V18 potential, which includes spin and angular momentum interactions between nucleons, to the five-dimensional space-time case.  This is the future perspective of the study.

\end{document}